\documentclass[
	a4paper, 
	10pt, 
	unnumberedsections, 
	twoside, 
]{LTJournalArticle}

\usepackage{amsmath,amssymb,amsfonts}
\usepackage{algorithmic}
\usepackage{xcolor}
\usepackage{url}
\urldef{\urlA}\url{https://github.com/tudtude/MICROSERVICE-RESTful-Nodejs} %
\urldef{\urlB}\url{https://github.com/dorsal-lab/tracecompass_nodejs\_support} 
\usepackage{array}
\usepackage[utf8]{inputenc}
\usepackage{fourier} 
\usepackage{array}
\usepackage{makecell}
\usepackage{cite}
\usepackage[numbers]{natbib}

\usepackage[caption=false,font=normalsize,labelfont=sf,textfont=sf]{subfig}
\usepackage{textcomp}
\usepackage{stfloats}
\usepackage{url}
\usepackage{verbatim}
\usepackage{graphicx}

\runninghead{Vnode: Low-overhead Transparent Tracing of Node.js-based Microservice Architectures} 

\title{Vnode: Low-overhead Transparent Tracing of Node.js-based Microservice Architectures} 

\author{%
	Herve M. Kabamba\textsuperscript{1}, Matthew Khouzam\textsuperscript{2} and Michel Dagenais\textsuperscript{1}\thanks{Corresponding author: \href{mailto:herve.kamba-mbikayi@polymtl.ca}{herve.kamba-mbikayi@polymtl.ca}\\ \textbf{Received:} November 17, 2023 }
}
\date{\footnotesize\textsuperscript{\textbf{1}}Computer and Software Engineering Department, Polytechnique Montréal\\ \textsuperscript{\textbf{2}}Ericsson Canada\\ }
\renewcommand{\maketitlehookd}{%
	\begin{abstract}
		\noindent Tracing serves as a key method for evaluating the performance of microservices-based architectures, which are renowned for their scalability, resource efficiency, and high availability. Despite their advantages, these architectures often pose unique debugging challenges that necessitate trade-offs, including the burden of instrumentation overhead.
With Node.js emerging as a leading development environment, recognized for its rapidly growing ecosystem, there is a pressing need for innovative approaches that reduce the telemetry data collection efforts, and the overhead incurred by the environment instrumentation. In response, we introduce a new approach designed for transparent tracing and seamless deployment of microservices in cloud settings. This approach is centered around our newly developed Internal Transparent Tracing and Context Reconstruction (ITTCR) algorithm. ITTCR is adept at correlating internal metrics from various distributed trace files, to reconstruct the intricate execution contexts of microservices operating in a Node.js environment.
Our method achieves transparency by directly instrumenting the Node.js virtual machine, enabling the collection and analysis of trace events in a transparent manner. This process facilitates the creation of visualization tools, enhancing the understanding and analysis of microservice performance in cloud environments.
	\end{abstract}
}

\begin{document}
\bibliographystyle{plainnat} 
\maketitle 


\section{Introduction}
\label{sec:introduction}
The swift advancement of technology has propelled the widespread adoption of microservice architectures, which highlights key aspects like availability, resilience, fault tolerance, and enhanced collaboration among teams. In such an architecture, each component operates independently and communicates through efficient, lightweight protocols. These architectures are highly favored due to their facilitation of collaborative efforts and their capability to meet modern challenges in application design, development, maintenance, and deployment \citep{newman2015building}. They bolster the system resilience, effectively manage failures, and seamlessly adapt to scaling requirements.
Nonetheless, these advantages come with certain challenges, particularly due to the heterogeneity of the components, issues in service allocation, and notably, concerns regarding the overall system performance \citep{lewis2014microservices}. 

Debugging issues in microservices architectures poses significant difficulties. System malfunctions can compromise user experience, and identifying the root cause of these issues can be elusive based on the information at hand. Even with error indications from protocols headers like HTTP,  the process of issue diagnosis can be challenging. Additionally, the arrangement of the components in a way that maintains system attributes such as availability and low latency raise intricacy.

The challenges of debugging \citep{sharma2015hansel, sambasivan2011diagnosing, sigelman2010dapper, tak2009vpath} and component arrangement \citep{sampaio2017supporting}, have been tackled through the implementation of distributed tracing techniques \citep{fonseca2007x, kaldor2017canopy, mace2015pivot}. However, applying these methods necessitates instrumenting the application source code, which brings additional overhead and the risk of altering the application behavior. Other strategies  extend instrumentation to dependencies to minimize changes at the application level. 

Though overhead issues have been partly addressed through sampling methods, both approaches to instrumentation present challenges. Instrumenting at the dependency level may fail to capture internal application logic issues like bugs, while instrumenting the application source code could potentially alter its functioning. Both methods entail modifications to the application, necessitating human effort and incurring extra costs.

The burden of instrumentation efforts has led to the development of strategies aimed at reducing these costs across several research domains \citep{luk2005pin, wang2008scisim}. In the microservices context, this issue has recently been tackled by advocating for the use of proxies, \citet{santana2019transparent},  as an intermediary layer for transparent tracing. However, while this spares the application from source code modification, it shifts the burden to setting up and configuring the proxies. 

Additionally, their operation involves intercepting system calls to insert trace context into requests, which can be problematic in public clouds where kernel access might be restricted.
We propose an innovative approach for tracing microservices in the Node.js environment, designed to overcome the limitations of existing methods while ensuring full transparency. 

This approach distinguishes itself in two key areas: (1) It obviates the need for developers to invest effort in establishing a collection infrastructure, and (2) It transparently analyzes collected traces, leading to visualization tools that map the interactions between microservices, thereby enabling the debugging of performance issues in a completely transparent manner.
The focus on Node.js stems from its complex, asynchronous environment. However, this approach could potentially be extended to other environments such as Java, Golang, and more.

The  remainder  of  the  paper  is  organized  as  follows:  Section~\ref{sec:basic} introduces the basic concepts of the study. Section~\ref{sec:related}  discusses  related  work  about  the  subject.   It is followed by Section~\ref{sec:solution} that presents the proposed approach for tracing and analysis. Section~\ref{sec:results} presents  the results of our work leveraging some  use  cases  to  show  our  its  pertinence  and relevance. An evaluation of our tool is conducted in Section~\ref{sec:evaluation}. In Section~\ref{sec:discuss}, we discuss the results obtained while in Section~\ref{sec:conclusion} a conclusion on the work is drawn.

\section{Basic concepts}
\label{sec:basic}
\subsection{Microservice architecture}
The microservices architecture is recognized as an application development strategy that involves breaking down applications into a series of loosely interconnected components. Its growing popularity can be attributed to the ease it brings to continuous delivery and its ability to enhance the scalability of applications [22].

Within this architectural framework, each component operates independently, managing a distinct function of the application. Communication between these components is facilitated through clearly established, lightweight protocols, typically using APIs. A key advantage of microservices architecture is the autonomy of its components, which grants development teams the flexibility to update and deploy individual services without impacting the overall application. This approach leads to a more streamlined development process and greater agility in software development [23].

While microservices offer numerous benefits, they also pose challenges in debugging. As an architecture comprising multiple heterogeneous components, performance monitoring tools must take this diversity into account. Typically, instrumentation is carried out on each component using tracing libraries specific to the component environment. To gain a comprehensive view of the system health, it is crucial to use distributed tracers. A benefit of microservice architectures is that the instrumentation phase can be treated as an application update, allowing for its flexible integration into the application deployment pipelines.

The health of microservices can also be monitored using logs, but the cost of this strategy becomes prohibitively high when the application consists of multiple nodes. The challenge lies not only in interpreting the inter-causalities among nodes but also in managing the large volume of data, which quickly becomes unmanageable. Tracing remains the best way to address this problem, as it allows for understanding the system operation as a whole and, if necessary, identifying bottlenecks.
\subsection{Distributed Tracing}
Distributed tracing is a strategy for collecting execution data from modern systems, particularly microservices architectures. It traces the lifecycle of a request as it passes through all the nodes in the system. Distributed tracing provides a hierarchical view of the trace, in which one can observe the time a request spends on each service \citep{sigelman2010dapper}.

Distributed tracing was introduced to address the complexities of distributed architectures, which are not suited to traditional debugging and tracing methods. In microservices environments, components need to interact to produce a result. In other words, when a request is issued from a particular node, it may need to interact with several other services in the infrastructure to return the result. In this context, when an issue arises along its path, it is necessary to identify where the bottleneck occurs to resolve the problem.

Traditional tracing techniques are not suitable because they are generally used for debugging monolithic applications. In the case of microservices, tracing involves injecting a trace context into the request headers to identify it throughout its lifecycle. In this way, the request can be properly aligned and hierarchized according to the service level it invokes over time. The various calls made by microservices during their interactions can thus be traced to understand the overall performance of the system \citep{sambasivan2011diagnosing}.

Distributed tracers such as Google Dapper \citep{sigelman2010dapper}, Zipkin, and Jaeger have revolutionized technology by offering the ability to collect data from each request, their execution times, and the causalities between services in the infrastructure.
Furthermore, analyzing the collected trace data can be complex. Distributed tracing systems generate a large amount of data, and interpreting them often requires advanced data analysis skills and a deep understanding of the system architecture \citep{mace2015pivot}.
Distributed tracing continues to evolve, with new improvements and integration into increasingly sophisticated tools and platforms. Therefore, it is necessary to propose tools that address shortcomings and bring more flexibility to the ecosystem.

\subsection{Node.js environment}
Node.js, an open-source environment, originates from the JavaScript V8 engine it is built on. It has brought the versatility of server-side JavaScript programming since its beginning in $2009$. Node.js revolutionized the web ecosystem by enabling developers to use JavaScript for server-side application development. 

Distinctive for its non-blocking, event-driven nature, Node.js is well-suited for high-performance applications. Its architecture is asynchronous and built on a single-thread model. In a Node.js application, a primary process known as the event loop orchestrates the execution of various events in different execution phases.

Tasks or events submitted for execution are first queued in a specific queue based on the event nature. The event loop traverses these phases, déqueues the events in each queue, and executes them. For blocking events, such as I/O operations, a thread pool is used to delegate execution and avoid blocking the event loop. Node.js is a multi-layered system. Internally, it comprises several components that work together to yield a result. 

However, at the lower layers of the operating system, such as in the kernel space, the Node.js process appears as a black box making system calls or generating context switch events. From this perspective, it is challenging to discriminate events being executed. A significant unresolved issue is linking high-level information, such as requests and invoked JavaScript functions, to actions performed in the V8 engine and the Libuv orchestration layer.

Current tools only allow visualization of high-level information, such as a request duration or the execution time of a service or JavaScript function. However, when such information is provided by distributed tracing tools, debugging is necessary to trace back to the cause.
Debugging such issues in Node.js is extremely challenging, since, as a single-thread system, all concurrent request executions are conflated into the same process. Distinguishing which request is causing performance issues is difficult because, even if a service shows high latency, it does not necessarily mean the service itself is at fault. For instance, the event loop might have been blocked by the execution of an non-optimized function, thus propagating the error to other pending requests.
Therefore, it's important to develop methods suitable to these environments. 

\subsection{Transparent Tracing of Node.js-based Microservices}
\label{sec:transparent_tracing}
A fundamental principle of the microservices architecture is the use of tools that are adequate for the task at hand. It enables the utilisation of various development environments, tools, and libraries. Due to the performance-restrictive architecture of microservices, effective and established methods are required to track and monitor their behaviour and, if necessary, address bugs. Distributed tracers are among the most efficient tools available to developers for observing and monitoring their systems. Through the use of collection modules, they are able to aggregate traces and enable their presentation through graphic user interfaces in the form of spans.

However, the adoption of these tools in the context of distributed systems, such as microservices architectures, necessitates a solid understanding of the source code of the application. Worse, modification of the latter is essential to incorporate tracepoints needed for information collection.
On the one hand, the application behaviour and programming structure are altered, and on the other hand, these tracers incur a substantial additional cost, which varies based on the execution environment and the technology employed.

Accelerating the performance analysis process requires methods that add little overhead on the system, and yet spare the developer from having to modify the program during the instrumentation phase. It simplifies the task by reducing the amount of work required and, in general, the associated costs.

The proposed method uses a transparent tracing approach that do not need any intervention from the practitioner, it uses the ITTR, which is a new technique we introduce that leverages the internal asynchronous mechanism of  Node.js for context reconstruction. In this way, request internal execution sequences can be reconstructed in a fully transparent manner.

\section{Related work}
\label{sec:related}
Recent research has demonstrated that distributed tracers are crucial for enabling the monitoring of interactions between microservices as studied by \citet{sampaio2017supporting}. This experience demonstrates the need for a more lightweight technique to trace such fine-grained architectures.

\citet{santana2019transparent}, suggested a novel transparent tracing methodology that leverages the kernel of the operating system to intercept system calls associated with communication among microservices. They proposed using a proxy that adds a neutral layer to the microservice, to intercept its interactions and correlate the information to deduce the causalities associated with the various requests. The interception of system calls ensures application tracing transparency, but the developer is responsible for configuring the infrastructure.

Statistically extracted dependency structures from documentation was used in service discovery by \citet{wassermann2011monere}, while fault detection was addressed by \citet{chen2002pinpoint}, through middleware instrumentation to log the respective components that process a particular request. A degree of transparency could be achieved, but such an approach requires the developer to dedicate much time to library instrumentation.

Distributed tracers Dapper and X-trace proposed respectively by \citet{sigelman2010dapper}, and \citet{fonseca2007x}, are able to trace the whole request lifecycle, expose its flow, and help diagnose issues throughout their execution. They rely on trace context injection mechanisms to reconstruct the context of the trace. A prior instrumentation phase of the application is required to activate the collection mechanism. In contrast to those approaches, our method does not inject the trace context into the request. It leverages the internal asynchronous mechanism of \texttt{Node.js} to reconstruct the request path.

Tracing request path strategies was also tackled by \citet{kitajima2017inferring} using heuristics. Request causality diagnosing algorithms were proposed by \citet{aguilera2003performance}. Both approaches offer a degree of transparency but rely on middleware instrumentation.

\citet{gan2019seer} introduced Seer, an online debugger designed to foresee quality of service (QoS) violations in cloud-based applications. Their research was conducted on a microservices framework, employing Memcached, which shares functional similarities with Redis as an in-memory database.

The process of debugging performance issues using Seer requires an instrumentation phase for microservices. This step is also implemented in the Memcached data store, particularly focusing on polling functions and various network interface queues. Seer has the ability to predict QoS violations using a model developed from deep learning techniques applied to upstream traces. By instrumenting the functions responsible for managing packet queuing at the data store level, Seer can effectively identify potential bottlenecks, especially those involving Memcached.
\section{Proposed solution}
\label{sec:solution}
\subsection{System Architecture}
\label{sec:architecture}
\begin{figure}[ht]
\centering
\includegraphics[width=0.9\linewidth]{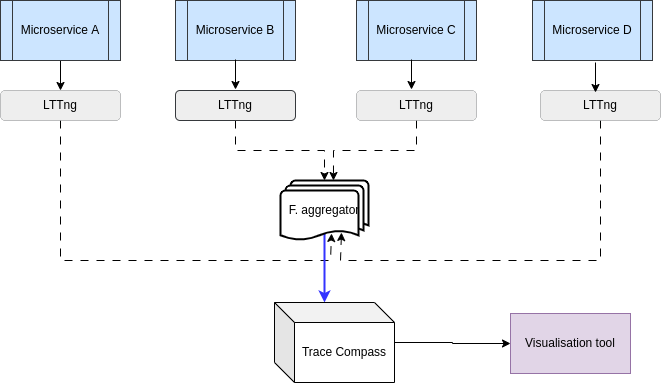} 
\caption{Example of a trace collection architecture used by our approach.}
\label{fig:architecture}
\end{figure}

The depicted diagram in Figure~\ref{fig:architecture} illustrates the operational framework of the system. Given that each microservice is deployed in a container, \texttt{LTTng} \citep{desnoyers2006lttng} (Linux Tracing Toolkits Next Generation) is enabled on each of them to capture traces. This produces several CTF (Common Trace File)-formatted local trace files. A file aggregator retrieves the aforementioned files for importing as an experiment under Trace Compass (TC) \citep{compass}, where analyses will be executed. The running of the analyses allows building an execution model based on the state system technology. It is constructed from the trace extracted attributes. 

The State History Tree (SHT) \citep{montplaisir2013state}, is a highly efficient data structure employed in the creation of the model, while reading the trace. It is used to store the attributes extracted from the trace, as well as various analysis-related data. TC provides the framework for modelling and developing such data structures. The optimisation technique employed enables the model to be queried in logarithmic time. It provides multiple features for the organisation of traces and the objectivity of analyses. This is achieved by employing multi-abstraction, highlighting, and filtering. TC enables the definition of the desired granularity of performance measurements and the application of analyses within the desired time limits.

The trace is collected by inserting tracepoints into the internal \texttt{Node.js} layers. LTTng is used to conduct static instrumentation. The tracepoints are initially inserted at the level of the \texttt{C++} bindings that interface with the native \texttt{JavaScript} modules, specifically at the functions that handle the socket communications in \texttt{Node.js}. Then, it is possible to extract the necessary information pertaining to the various sockets and their attributes for the correlation and context reconstruction stages.

Information on the request attributes (methods, addresses, and ports) is transmitted directly from native \texttt{JavaScript} modules to \texttt{C++} bindings. In this manner, the expense of parsing \texttt{HTTP} requests is avoided, and an \texttt{LTTng} probe is inserted at that juncture. The engine that generates asynchronous resource identification numbers is probed during the second instrumentation step. This technique allows accessing, at the origin, the identifiers of the execution contexts associated with asynchronous resources, as well as the identifiers of the execution contexts of the resources that generated them. In addition, it enables monitoring of their entire life-cycle, from creation to destruction.

Therefore, the costly \texttt{Node.js} \texttt{Async Hooks} API is unneeded for the monitoring of asynchronous resources. Tracing is performed directly within the VM, which significantly reduces the overhead in this context. \texttt{LTTng} has become known as the fastest tracer in the world and incurs minimal system overhead. It permits the creation and collection of events, which are then loaded into TC. Developed extensions in the latter facilitate the creation of event analyzers and handlers. It is a free, open-source tool that allows for the analysis of traces and logs. The extensibility of the system enables the creation of graphs and views, as well as the extraction of metrics.

\subsection{Patterns-based context reconstruction formalization }
\label{sec:pattern_base_reconstruction}
When conducting the analysis using TC, a state system is built. The states encompass all the events that are deemed acceptable by the analyzer for the purpose of instantiating and activation of the different system transitions. Our approach for reconstructing the execution contexts of all the requests relies on identifying specific sequences of system transitions. These sequences are retrieved from the global state system and serve as the detected patterns. By traversing the trace to construct the state system, concurrent state subsystems belonging to the global system are identified and correlated with the concurrent requests to which they are bound.
The obtained state system can be regarded as a finite state machine consisting of six components.\\\\
Let $M$ be a finite state machine:
\begin{equation}
	M = <P, F, B, G, s, tr> 
	\label{eq:1}
\end{equation}
where:\\
$P$ represents the state space\\
$F$ represents the event space\\
$B$ represents the action space \\system's
$G$ represents a subset of P \\
$s$ represents an element of P, the initial state\\
$tr:P\times F\xrightarrow[]{}P\times A$, represents the state transition function\\

\begin{equation}
\label{eq:tr_state}
tr(p, f) = (q, b)
\end{equation}
\begin{equation}
\label{eq:tr_ext}
TR:E^* \xrightarrow[]{}P^*\times B^*
\end{equation}
where $X^*$ represents all the sequences of members belonging to X. In other words, the patterns that model the concurrent requests execution in the state system are sequences of members of the global state system.

The transition function can be extended as defined in Equation~\ref{eq:tr_state}. In this case, if $p$ is the active state of M, and if there is an occurrence of event $f$, then $q$ becomes the new active state of the system, therefore action $b$ is taken. The handling of events during the execution of the analysis make the system transition to multiple states. For each state, related actions are undertaken.\\

\begin{equation}
\label{eq:01}
tr(s, p) = (k, a)
\end{equation}
\begin{equation}
\label{eq:02}
tr(k, t) = (w, b)
\end{equation}
\begin{equation}
\label{eq:03}
tr(w, h) = (n, c)
\end{equation}
\begin{equation}
\label{eq:fin}
TR(pth) = (skwn, abc)
\end{equation}
Consider the events $p$, $t$, $h$, the states $s$, $k$, $w$ and the actions $a$, $b$, $c$ defined in Equation~\ref{eq:01}, Equation~\ref{eq:02} and Equation~\ref{eq:03}. Then \texttt{pth} is an event sequence in $F^*$. As defined by Equation~\ref{eq:fin}, the systems should transition from its initial state, $s$, to $k$, then to $w$ and $n$. For each transition, the system will perform the actions $a$,$b$ and $c$.
The inputs to the model are the different accepted events from the trace while the handler is running. Actions are taken for each of the accepted events to construct the global state system and build the SHT. 
The patterns are modelled as subset of the global state system identifying transition sequences.

\subsection{Transparent trace collection}
\label{sec:context_recon}
Microservice interactions occur through message passing across network sockets. In microservices using HTTP as the communication protocol, when one microservice wants to send a message to another, it first establishes a connection with the remote component.
Internally in Node.js, native modules managing network sockets communicate with the Node.js virtual machine to request information about the created socket context.
In a single-threaded environment like Node.js, multiple events are managed concurrently. Node.js implements an internal mechanism to track different asynchronous resources by assigning them an identifier.

For instance, if a function with identifier "12" first creates a socket, this socket might receive identifier "13," implying that this new resource was created in the "12" context. Similarly, any new resource initiated within the socket's context will have "13" as its execution context, aside from its own resource identification number.
Managing asynchronous resources in environments like Node.js is extremely complex and challenging. Async Hooks, an experimental API, was proposed to track the lifecycle of these asynchronous resources.
However, this API is significantly costly in terms of system performance, sometimes adding overhead by up to 50\%.

For most production applications, this overhead is not tolerable. This is because each call to the API requires crossing the JavaScript/C++ barrier of the V8 engine, which incurs a very high cost to the system.
The proposed solution circumvents this by addressing and collecting information directly at the source, at the level of the engine responsible for generating asynchronous resources.
In other words, our solution involves definitively instrumenting the Node.js virtual machine to transparently collect network communication data.

\begin{figure}[ht]
\centering
\includegraphics[width=0.8\linewidth]{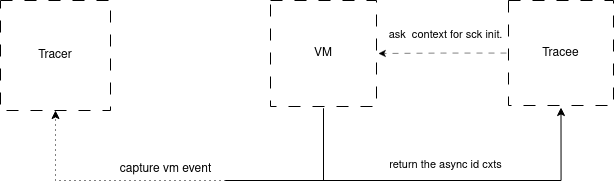} 
\caption{Internal resource context  creation by the VM}
\label{fig:context_cre}
\end{figure}

Figure~\ref{fig:context_cre} shows the internal process of execution context creation at the VM level when a network connection is established. It can be observed that when a microservice sends a message to another component, Node.js native network modules request the context from the asynchronous resource manager. 

This manager initializes the new resource and assigns it a generated number. The number assigned to the created socket allows for the identification of the execution context of objects. For example, when a response to a request returns, Node.js checks the socket number (context) in the request header to route it to the waiting resource. This is Node.js multiplexing function, given its single-threaded nature.

To trace interactions between microservices, five main tracepoints were inserted into the native module and the Node.js virtual machine: \texttt{http\_server\_request}, \texttt{http\_server\_response}, \texttt{http\_client\_request}, \texttt{http\_client\_response}, and \texttt{async\_context}.

The \texttt{http\_server\_request} tracepoint is activated when the server receives a request to process, such as when a microservice receives a specific request. The \texttt{http\_server\_response} tracepoint is activated when the server returns a response after processing the request. The \texttt{http\_client\_request} tracepoint is activated when a component emits a request, for example, when a microservice contacts another microservice. The \texttt{http\_client\_response} tracepoint is activated when the response is returned to the sender.

\begin{figure}[ht]
\centering
\includegraphics[width=0.8\linewidth]{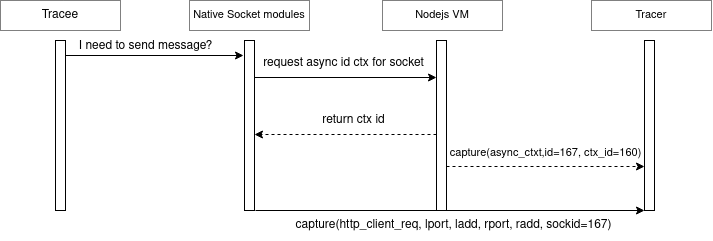} 
\caption{Internal communication process when sending a request. Tracepoints are activated}
\label{fig:context_send}
\end{figure}

Figure~\ref{fig:context_send}  shows how this process unfolds when a request is issued by a microservice. First, native network functions get the socket context information, containing the socket number and the number of the resource that created it, to preserve hierarchy and execution sequence. The creation and initialization of the new resource by the VM trigger an event captured by LTTng (the event is recorded as \texttt{async\_context}). 

This event exports various information in its attributes, including the identification number of the created resource, the identification number of the parent resource, and the resource type. After the request is sent by the native network functions, the \texttt{http\_client\_request} tracepoint is activated and captured by \texttt{LTTng}. It is exported with the attributes represented in Figure~\ref{fig:context_cre}.

\begin{figure}[ht]
\centering
\includegraphics[width=0.8\linewidth]{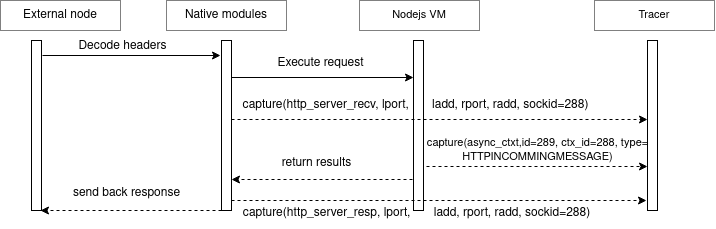} 
\caption{Processing incoming request. Tracepoints get triggered.}
\label{fig:context_inc}
\end{figure}

When the destination microservice receives the request, it can be observed in Figure~\ref{fig:context_inc} that the request is first decoded by Node.js network functions, and at the same time, the \texttt{http\_server\_receive} tracepoint is activated. The event is exported to LTTng with all the attributes present in Figure~\ref{fig:context_inc}. At the end of the request processing, the request is returned, activating the \texttt{http\_server\_response} tracepoint, which is also exported with all its attributes.

Activating the different tracepoints produces a file in the CTF format at the end of the system tracing. This file containing the transparently collected information can be very large depending on several factors, such as the system load or the duration of its tracing. Therefore, it is crucial to employ automated methods to analyze the trace to extract relevant information. The next section addresses the new analysis approach we propose in this work.
\begin{figure}[ht]
\centering
\includegraphics[width=0.8\linewidth]{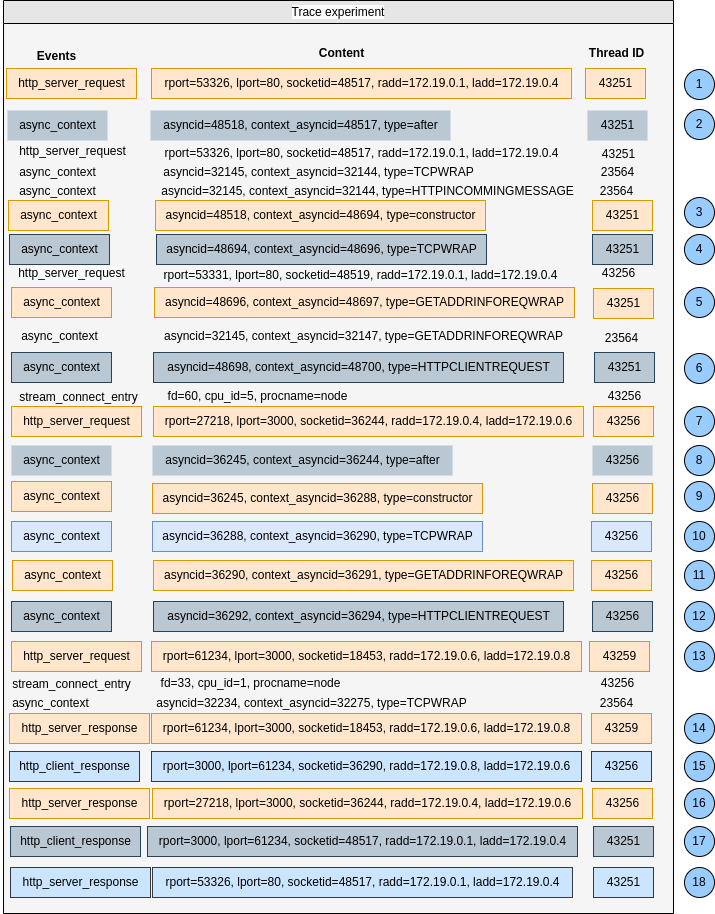} 
\caption{Example of a user trace experiment. Several trace files are aggregated and opened as experiment in TC}
\label{fig:trace}
\end{figure}

\begin{figure}[ht]
\centering
\includegraphics[width=0.8\linewidth]{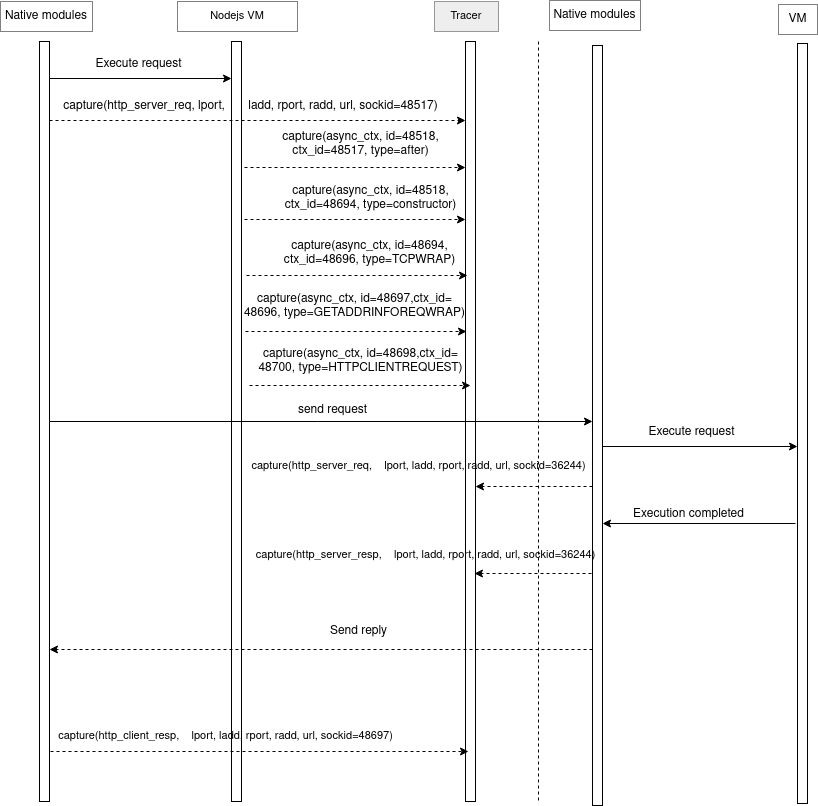} 
\caption{Communication between two microservices.}
\label{fig:context_two}
\end{figure}

\subsection{Trace analysis technique}

Performance debugging in microservices involves a phase of trace collection that necessitates system instrumentation. This step is crucial for gathering the necessary data to interpret the system operation. Analysis abstracts the system overall functioning to avoid delving into the minutiae of the data extracted from the system.
However, effective, robust, and rapid methods are required to utilize the data collected in the initial phase, as the files can become exceedingly large, containing millions or even billions of events. This necessitates the use of tools capable of handling such vast quantities.
Our approach conceptualizes system executions as finite state machines. A request is viewed as an automaton transitioning through states based on trigger events. The automaton sequence of transitions represents a series of events occurring during system execution.

Our method achieves its transparency in analysis by matching pattern sequences observed during request executions. An initial understanding of system activity based on collected traces allowed the identification of recurring patterns in microservice interactions within Node.js. We then modeled these patterns as finite state machines to understand the system various states to debug its performance.
We utilized TC to load the trace for our analyses. TC' extensibility enables the development of visualizations based on analyses, providing essential tools for system performance analysis.
To preserve the automaton different states, we used a particular, expandable data structure optimized for supporting very large file sizes, known as the SHT.

Observing Figure~\ref{fig:context_two}, a request arrives at the gateway microservice, which must redirect it to the relevant service microservice.

In this case, activating the \texttt{http\_server\_request} tracepoint initiates the state machine and sets it to the “receive request” state. To preserve this state, the highest level in the hierarchy, the automaton state is recorded in the SHT with data extracted from its attributes. The automaton transitions to the next state when the \texttt{async\_context} event is encountered in the trace, and its attribute type is “after”. Here, the sockid value of the previous \texttt{http\_server\_request} event is matched with the \texttt{ctx\_id} attribute of the \texttt{async\_context} event, ensuring that the current event occurred within the context of the ongoing request. 

The next state is activated when the \texttt{async\_context} event’s attribute type is “constructor”. At this level, the id attribute value is matched with that of the previous state (48518 and 48518), ensuring that the reconstructed sequence is linked to the initial request. The automaton transitions to the next state when the \texttt{async\_context} event type value equals “TCPWRAP”. 

To maintain the sequence context, the id attribute is matched with the \texttt{ctx\_id} of the previous state (48694 and 48694). The next automaton transition occurs when the \texttt{async\_context} event type value is “GETADDRINFOREQWRAP”, with its \texttt{ctx\_id} matched to the previous state (48696 and 48696). Finally, the system transitions to the next state when the \texttt{async\_context} event type value is “HTTPCLIENTREQUEST”. Here, the id attribute value must be one order higher than the previous state (48698 and 48697).

At this point, the asynchronous sequences through which the request received by the gateway microservice passes to be sent to the concerned microservice are transparently reconstructed. This state sequence is the model followed by Node.js to communicate with microservices via HTTP (REST API). The type attribute values are obtained directly from the Node.js virtual machine, representing the different asynchronous resources created during request execution.
Each system state is recorded with its attribute values in the SHT. The context allows hierarchically inserting states and attributes into the SHT tree to form a hierarchy defining the sequence of automaton execution, their start and end, thus enabling visualizations to extract information for studying the system performance.
The outcome of the algorithm yields a structured and hierarchical representation of the diverse interactions across microservices, achieved through a fully transparent process.

\section{Results}
\label{sec:results}
In this section, we demonstrate the capabilities of our tool using three use cases scenario. The objective is to effectively articulate the anticipated outcomes derived from the utilisation of \texttt{Vnode}. The \texttt{Nodejs-Restful-Microservices} \footnote{\urlA} application is utilised for this purpose.

\subsection{use case 1}
\begin{figure*}[ht]
\centering{\includegraphics[width=14cm,height=2cm]{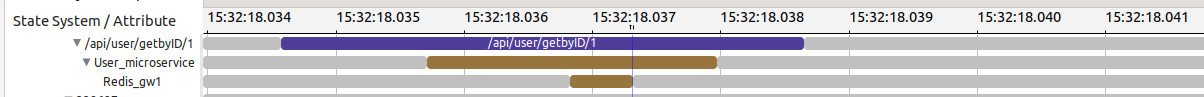}} 
\caption{Vertical Span Model (VSM) representation. The label “A” depicts a span as it appears in distributed
tracers. The label “B” depicts a vertical span representing the flow of the request sequences in all layers.}
\label{fig:usecase1}
\end{figure*}

In the initial use case, simultaneous requests are performed in order to retrieve specific information pertaining to an individual user. The requests made are of the \texttt{GET} type. Executing those requests after deploying the application generates multiple CTF-formatted trace files. As described previously, they are aggregated and imported as an experiment in TC. Figure~\ref{fig:usecase1} depicts the visual outcome of our analyses. The reconstruction of the alignment of request execution flows in accordance with their respective contexts can be observed. 

After receiving the request, the server transmits it to the user microservice. The request is then forwarded to the Redis gateway since the data has been put into memory in the Redis data structure. The \texttt{Vnode} facilitates the transparent horizontal sequencing of requests and can be seamlessly integrated into the application development and operation pipelines.

\subsection{use case 2}
\begin{figure*}[ht]
\centering{\includegraphics[width=14cm,height=5cm]{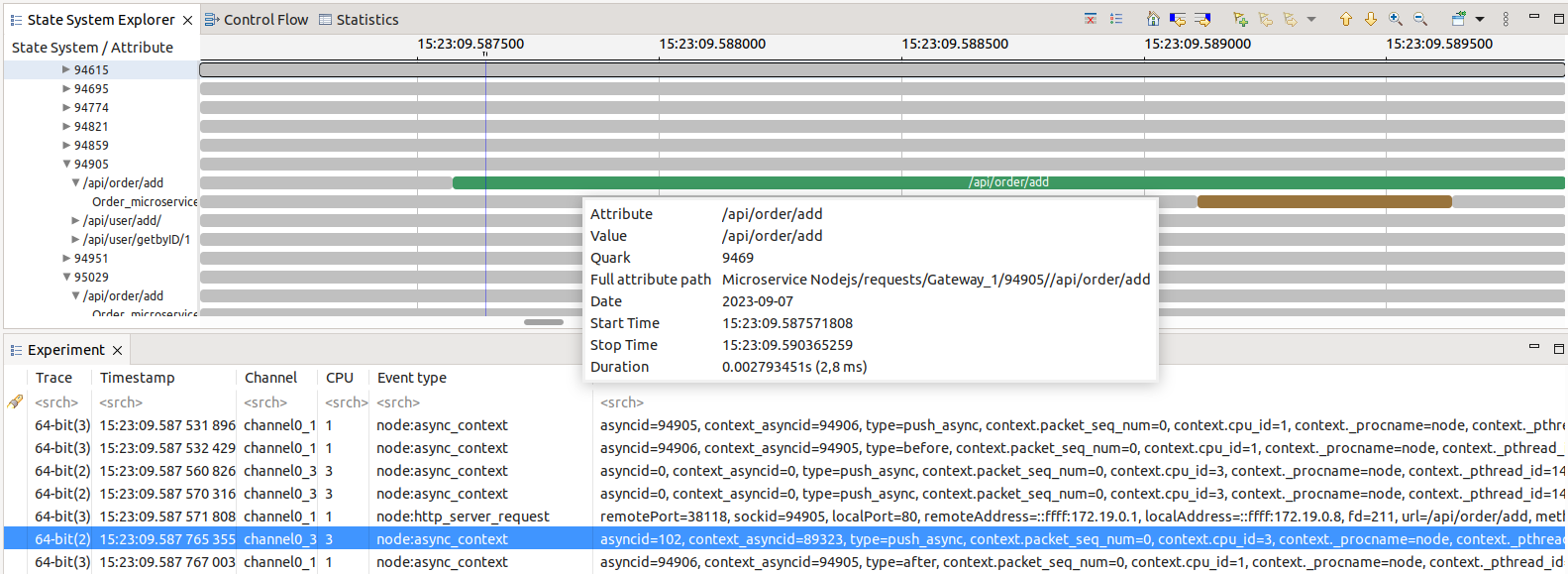}} 
\caption{Vertical Span Model (VSM) representation. The label “A” depicts a span as it appears in distributed
tracers. The label “B” depicts a vertical span representing the flow of the request sequences in all layers.}
\label{fig:usecase2}
\end{figure*}

The second use case involves the sending of \texttt{POST} requests to place item orders. Figure~\ref{fig:usecase2} demonstrates the capability of \texttt{VNode} to smoothly rebuild request execution contexts. it depicts the output of the algorithm execution. In contrast to the initial use case, upon receipt of the request by the server, it is promptly forwarded to the microservice responsible for handling orders. As the operation entails the insertion of data to the database, the microservice executes the operation directly, bypassing the \texttt{Redis} gateway.

By analysing the two use cases, the unique feature of \texttt{Vnode} regarding the reconstruction of the communication architecture becomes evident. The strength of \texttt{Vnode} resides in its capacity to enable developers to comprehend and visualise the communication architecture of microservices systems implemented in \texttt{Node.js}. The developer does not need to grasp the application code or inner workings to comprehend how its components interact internally. \texttt{Vnode} reconstructs each API call execution sequence transparently and presents the result in a visual and interactive tool.

\subsection{use case 3}
\begin{figure*}[ht]
\centering{\includegraphics[width=14.4cm,height=5cm]{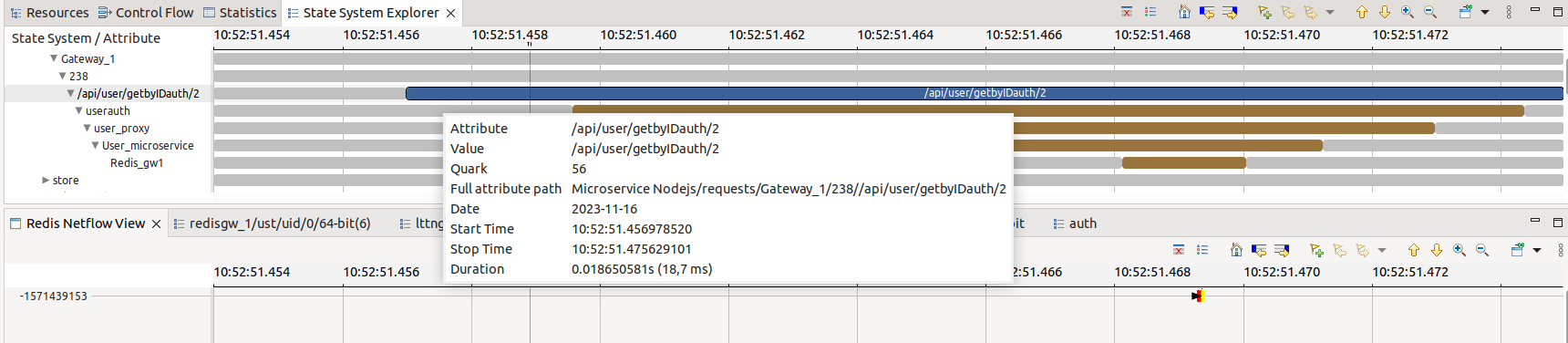}} 
\caption{Request sent through the authentification service}
\label{fig:redis1}
\end{figure*}

In this third scenario, requests are sent to a user authentication service to obtain the tokens necessary for user session validity. As shown in Figure~\ref{fig:redis1}, in this case, once the request reaches the gateway microservice, it is automatically redirected to the authentication microservice. After processing the request, the latter redirects it to the microservice acting as a proxy for the user service, before being redirected again to the user microservice, which retrieves user information from the Redis database through the redis gateway microservice.

\begin{figure*}[ht]
\centering{\includegraphics[width=14cm,height=5cm]{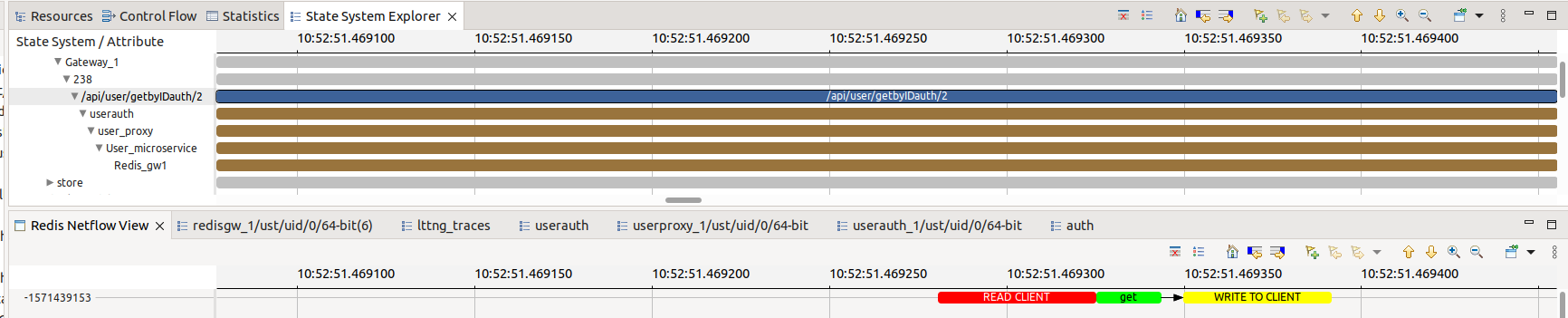}} 
\caption{Zooming in into the Redis analysis that is joined to the Nodejs one.}
\label{fig:redis2}
\end{figure*}

Aligning the spans allows the complete visualisation of the time the request spent in each microservice. By combining this analysis with the one presented  the \texttt{Redis} one, shown on the bottom in Figure~\ref{fig:redis1} and Figure~\ref{fig:redis2}, we can observe that the execution of the "get" command on the Redis server took an extremely short time, only $40$ microseconds. However, the request took at least $18$ milliseconds to complete. It is clear that the majority of the time was spent during the interactions between the microservices.

These results demonstrate a trace can be collected and analyzed in a completely transparent manner without requiring developer intervention. To the best of our knowledge, there is currently no approach that allows for this. In this context, the execution flow and operational architecture of the microservices can be visualized without prior knowledge of the system's implementation.

\section{Evaluation}
\label{sec:evaluation}
This section presents an evaluation of our tracing approach. Experiments we carried out in different scenarios for the purpose of validating our strategy and comparison to state of the art approaches.
\subsection{Objectives}
A comprehensive assessment was conducted to evaluate the performance of our solution with respect to the incremental cost incurred by the system being evaluated. The process encompasses three primary parts: i) an evaluation of the overall overhead incurred by the implementation of our tool; ii) a comparison of the overhead associated with our tool in relation to other tracing methodologies; and iii) the assessment of metrics that directly influence the development and operation of microservices.

\begin{table}
  \begin{center}
  \caption{\label{tab:Table1}Defined experiment parameters}
    \begin{tabular}{ | c | c |}
      \hline
      \thead{Parameters} & \thead{value} \\
      \hline
      number of executed requests &  \makecell{1200}\\
      \hline
      Time between requests &  \makecell{Randomly generated \\ (Gaussian distribution)}\\
      \hline
      Tracing configurations &  \makecell{No Tracing \\ State-of-the-art Rbinder \\ Our strategy}\\
      \hline
    \end{tabular}
       
  \end{center}
\end{table}
\subsection{Experiments}
\begin{figure*}%
    \centering
    \subfloat[\centering Averaged Response time for microservice operation "user get"]{{\includegraphics[width=6cm, height=3cm]{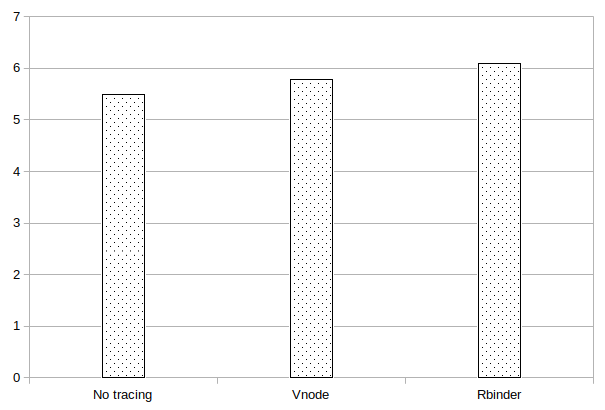} }}%
    \qquad
    \subfloat[\centering Averaged Response time for microservice operation "user add".]{{\includegraphics[width=6cm, height=3cm]{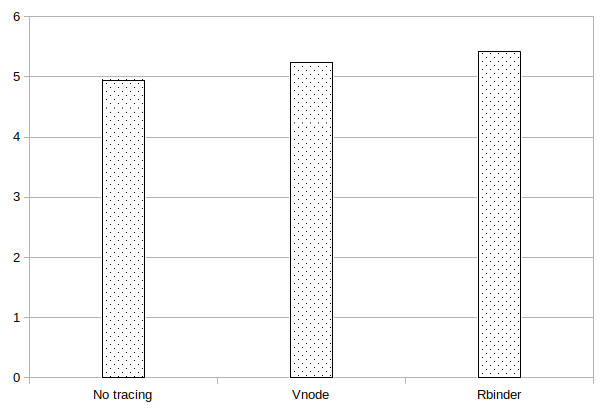} }}%
    \qquad
    \subfloat[\centering RAM usage in no tracing and Vnode scenarios.]{{\includegraphics[width=6cm, height=3cm]{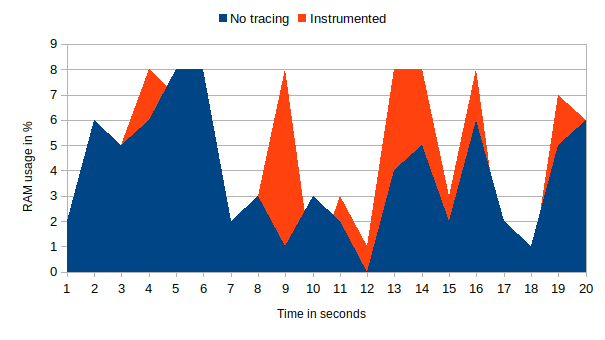} }}%
    \qquad
    \subfloat[\centering CPU usage in no tracing and Vnode scenarios.]{{\includegraphics[width=6cm, height=3cm]{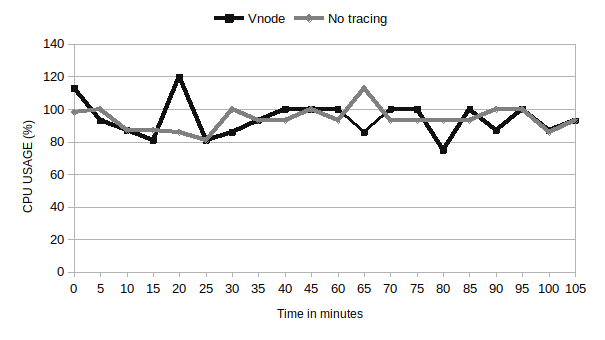} }}%
     \qquad
    \subfloat[\centering Trace size versus analysis time.]{{\includegraphics[width=6cm, height=3cm]{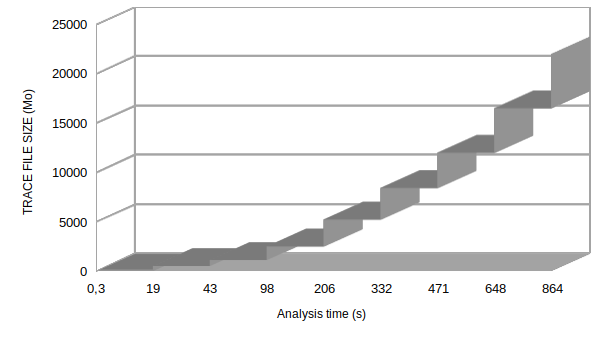} }}%
     \qquad
    \subfloat[\centering CPU usage when running analyses.]{{\includegraphics[width=6cm, height=3cm]{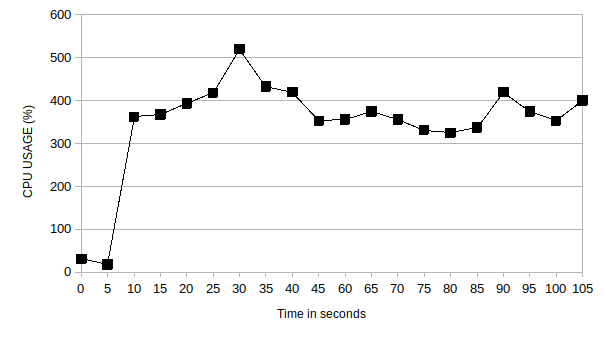} }}%
     \qquad
    \subfloat[\centering RAM usage when running analyses.]{{\includegraphics[width=6cm, height=3cm]{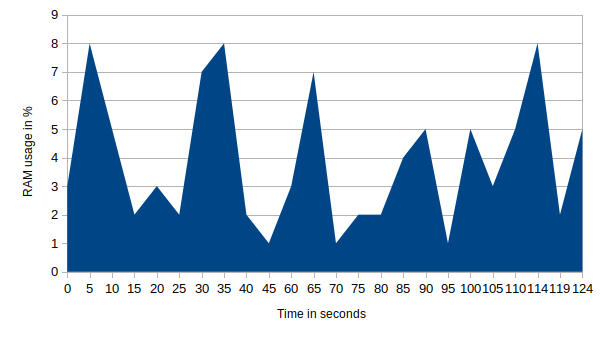} }}%
    \caption{Experiment results and impact on physical resources}%
    \label{fig:experiments}%
\end{figure*}
We utilised the \texttt{Node.js} microservice application to generate the user traffic for the experiments. Response time was selected as the metric for evaluation because it accurately reflects the user experience as highlighted by \citet{menasce2002qos}. The parameters employed in our investigations are presented in Table~\ref{tab:Table1}. \texttt{Apache Jmeter}\citep{jmeter} was utilised to generate $1200$ \texttt{HTTP POST} requests directed at the "add order operation" and $1200$ \texttt{HTTP GET} requests directed at the "get user operation" the two operations with the longest critical paths.

We considered three evaluation scenarios when carrying out our experiments. The evaluation of the application was done without the use of any instrumentation. In this particular scenario, the microservices were deployed with their original \texttt{node.js} versions without any alterations. The average response time was observed. \texttt{JMeter} is employed for the purpose of generating user load. In the second scenario, the application is deployed using the strategy proposed by \citet{santana2019transparent}. The average response time was also observed. In the final scenario, the instrumented \texttt{Docker} images for \texttt{Node.js} are deployed alongside the application. The observation of response times was also performed.

In order to evaluate the effectiveness of the Rbinder technique\citep{santana2019transparent}, we implemented it in conjunction with our microservice application and proceeded to configure the different proxies accordingly. No system call activation was performed in our strategy; only \texttt{Node.js} Docker images were deployed alongside the application. The technique implemented does not necessitate changes to the application deployment procedure. One of the benefits of this approach is its ability to provide transparency in both deployment and tracing processes. The analyses are characterised by transparency, as they generate visual results without any involvement from the developer.

The process requires starting the \texttt{LTTng} tracer on every container to capture the application trace, afterwards halting the tracer, and automatically aggregating the traces. The microservices are then imported onto the TC platform, where an analysis is conducted and a visualisation depicting the interactions between these microservices is generated.

\section{Discussion}
\label{sec:discuss}
The outcomes of our studies are depicted in Figure~\ref{fig:experiments}(a). We can observe that the response time of the application executed using our method is comparable to that of the uninstrumented application. The response time for the "get user" operation is 0.0055s when no instrumentation is conducted, 0.0058s when our strategy (Vnode) is employed with instrumented versions of Node.js, and 0.0061s when Rbinder is used. Analysing the additional cost caused by our approach reveals that it is $1.054$ time slower than the untraced application, compared to Rbinder, which has a response time of $1.095$ slower.
It can be concluded that the impact of tracing is quite acceptable and that response times are comparable to those of the application without tracing.

In the second scenario, experiments were conducted by initiating \texttt{POST} requests to the "user add" operation. The obtained response times are depicted in Figure~\ref{fig:experiments}(b). It is clear that the supplementary cost associated with tracing, on average, is consistently of a comparable magnitude as in the preceding scenario. The mean response time for the untraced program is $0.00494$ seconds. For the Rbinder application, the average response time is $0.00542$ second, and for the Vnode application, it is around $0.00524$ second. The examination of overhead resulting from tracing with our approach reveals that it is comparable to that of state-of-the-art approaches.

Figure~\ref{fig:experiments}(c) illustrates the central processing unit (CPU) utilisation in the scenario when the application is executed without any tracing. Upon comparing the aforementioned data with Figure~\ref{fig:experiments}(d), which illustrates the CPU utilisation while employing our proposed methodology, a marginal increase in resource use is observed. 

The LTTng tracer has a minimal impact on the host system, and this effect is diminished when tracing is performed at the user level with a selection of events. The outcomes of our studies are illustrated. It is evident that the response time of the application implemented using our approach closely approximates the response time of the non-instrumented application. Through a thorough examination of overhead incurred by our methodology, it was been determined that it exhibits a performance slowdown of $1.054$ times when compared to the untraced application. This is in contrast to Rbinder, which demonstrates a response time of $1.095$.
It can be concluded that the impact of tracing is acceptable and that response times are comparable to those of the application without tracing.

Figure~\ref{fig:experiments}(e) and (f) respectively depicts the CPU utilisation and time to run the analysis according to the trace sizes. For different size of the trace, the needed time shown. The experiments show that our algorithms run in acceptable times.
\section{Conclusion}
\label{sec:conclusion}
This work presented a new method for tracing \texttt{Node.js} microservice architectures. It emphasized the importance of tracing transparency so as to reduce the time spent on the performance analysis and validation phases of microservices. By adopting a tracing paradigm based on context reconstruction through \texttt{Node.js} virtual machine instrumentation, a specific algorithm has been developed for various multi-layer metrics correlations.
In this case, the burden imposed by the instrumentation phase to developers can be avoided. The  presented approach not only allows the transparent tracing of microservices, but it also provides a framework for uncovering the exécution architecture of microservice. 

Our approach could be improved by extending it to the support of other communication protocols, such as websockets.

Another  potential avenue for further exploration involves doing an in-depth investigation into the root cause of performance issues, which our approach enables to pinpoint for the purpose of conducting a root cause analysis.

 \bibliography{cas-refs}


\end{document}